\documentclass[10pt,a4paper,twocolumn]{article}

\usepackage[margin=2cm]{geometry}
\usepackage[T1]{fontenc}
\usepackage{amsmath,amssymb}
\usepackage{graphicx}
\usepackage{cite}
\usepackage{authblk}
\usepackage[hidelinks]{hyperref}

\title{\textbf{Contrast-Enhanced Nonlinear Imaging
with Moir\'e Flat-Band Superlattices}}

\author[1]{Amin Zamani}
\author[2]{Lihang Feng}
\author[1]{Jianji Liu}
\author[1]{Ze Zheng}
\author[1]{Gabriel Sanderson}
\author[1]{Mohsen Rahmani}
\author[2,*]{Peilong Hong}
\author[1,$\ddagger$]{Lei Xu}

\affil[1]{Advanced Optics \& Photonics Lab,
Department of Engineering, School of Science and Technology,
Nottingham Trent University, Nottingham NG11 8NS, United Kingdom}

\affil[2]{School of Physics and Astronomy,
Anqing Normal University, Anqing, Anhui 246133, China}

\affil[*]{\href{mailto:p.l.hong@aqnu.edu.cn}
{p.l.hong@aqnu.edu.cn}}
\affil[$\ddagger$]{\href{mailto:lei.xu@ntu.ac.uk}
{lei.xu@ntu.ac.uk}}

\date{}

\begin{document}

\twocolumn[
\begin{@twocolumnfalse}
\maketitle
\begin{abstract}
Nonlinear metasurfaces offer a compact platform for infrared-to-visible conversion and imaging, yet resonant enhancement is often accompanied by restricted angular acceptance, limiting the range of transverse spatial frequencies that can interact efficiently with the nonlinear modes. Here, we address this challenge using a one-dimensional photonic moir\'e superlattice supporting multiple quasi-flat bands in the near-infrared. These moir\'e modes combine strong near-field enhancement with broad momentum-space accessibility, enabling efficient third-harmonic generation (THG) while maintaining stable excitation over a large angular range. Experimentally, we observe strong THG emission near $467~\mathrm{nm}$ under broadband excitation from $ 1180$ to $ 2000~\mathrm {nm} $. We further employ the nonlinear moir\'e superlattice for infrared imaging by converting an infrared target into visible through the THG process. By encoding image information into displacement-dependent THG signals and decoding their spatial variations, we experimentally resolve weak features and internal textures with improved contrast compared with THG imaging using nonlinear films or conventional nonlinear metasurfaces. These results establish photonic moir\'e superlattices as a versatile platform combining efficient frequency conversion, broad angular acceptance, and mode-sensitive spatial encoding for high-contrast, label-free nonlinear imaging with enhanced feature discrimination.
\end{abstract}
\vspace{1em}
\end{@twocolumnfalse}
]

\section{INTRODUCTION}
Nonlinear optical imaging techniques use intensity-dependent light-matter interactions to transfer spatial information from an object to newly generated frequencies, offering a powerful platform to convert infrared signals into the visible spectral range and detect them using low-cost silicon cameras~\cite{krishnan1978infrared,barh2019parametric,fang2026mid}. The separation of illumination and detection wavelengths allows the two channels to be independently optimized to improve imaging performance and enable additional imaging functionalities~\cite{brady2009optical,hu2024metasurface}. In recent years, resonant metasystems have emerged as compact platforms to facilitate nonlinear imaging, offering control over the spatial, spectral, and polarization properties of nonlinear emission to allow tailored imaging functionalities~\cite{rahmani2018nonlinear,huang2023resonant,schulz2024roadmap}. Resonance-induced enhancement of local electromagnetic near fields can further increase the efficiency of second- and third-order nonlinear optical processes~\cite{camacho2021infrared,valencia2024enhanced,sanderson2024infrared,zheng2024broadband,zheng2026nonlinear,liu2026high,zamani2026linear}. 
However, a fundamental trade-off exists between nonlinear enhancement and imaging capability. Conventional resonant metasurfaces typically rely on one or a few isolated modes, for which strong field confinement is generally accompanied by narrow spectral bandwidths and limited angular acceptance~\cite{huang2023resonant,schulz_2024_21w38-xyx27,kuznetsov2024roadmap}. As a result, the resonance frequency varies with the in-plane wavevector, and only a restricted range of incident angles, and hence transverse spatial frequencies of the object, remains resonant at a fixed excitation wavelength. This angular selectivity not only limits the obtained conversion efficiency, but also limits the spatial information captured in the nonlinear image.
Photonic flat bands offer a promising route to overcome this trade-off by sustaining resonant enhancement over a broad range of transverse wavevectors. Their weak dispersion allows multiple spatial-frequency components of an object to couple resonantly at the same excitation frequency, while the associated low group velocities, enhanced optical density of states, and field localization can strengthen nonlinear light-matter interactions~\cite{Leykam2018Perspective,Danieli2024Flat}. These properties are expected to enable efficient nonlinear conversion across a wider range of spatial frequencies, improving image contrast while preserving finer spatial detail.

Photonic moir\'e superlattices provide a versatile platform for realizing and engineering multiple weakly dispersive photonic bands~\cite{du2023moire,Tang:26}. They can be formed by superimposing periodic structures with slightly different lattice constants or relative orientations, introducing a modulation on a length scale larger than the constituent periods. The enlarged moir\'e supercell folds the original constituent photonic bands into a reduced Brillouin zone. Although band folding alone does not flatten the dispersion, the moir\'e-induced coupling and hybridization between the constituent modes can reshape the folded bands and, under suitable conditions, produce weakly dispersive minibands~\cite{PRL2021Dong,LSA2021Tang,PRR2022Nguyen,Raun2023GaN,APN2023Hong}.
Compared with a conventional periodic metasurface designed around one or a few isolated resonances, moir\'e superlattices offer additional design freedom to engineer multiple weakly dispersive resonant channels with distinct spatial field profiles and momentum-space responses~\cite{APN2023Hong,hong2025hybrid}. Together, these modes enhance the local field and extend resonant coupling across a wider range of transverse wavevectors. This combination is especially advantageous for nonlinear imaging: the strong intensity dependence of harmonic generation amplifies small variations in excitation and local-field distribution, enhancing image contrast, while multiple quasi-flat bands allow a wider range of object spatial frequencies to contribute to image formation, potentially recovering finer features.

Despite extensive studies of light localization, diffraction, and optical nonlinear effects in photonic moir\'e systems~\cite{NP2020Fu,NN2021Mao,OL2022Hong,PRL2022Huang,luan2023reconfigurable,du2024nonlinear,qu2025beam}, their potential for nonlinear imaging remains largely unexplored. Recent studies have primarily focused on enhanced light–matter interactions and frequency conversion associated with flat-band states~\cite{wang2024experimental,wang2025moire,yan2025cavity,ji2025giantly,hu2026Flatband}. How multiple moir\'e-induced flat bands shape the angular acceptance and spatial-frequency response of nonlinear conversion, and how these properties can be harnessed for image formation, remain open questions. 

In this work, we harness multiple photonic flatbands supported by a moir\'e superlattice to enhance the THG for nonlinear imaging. Beyond resonant enhancement, we employ the spatially varying mode profiles within the moir\'e supercell with lateral translation changes their overalap with the projected traget field, encoding the spatial information into the displacement-dependent THG emission. These nonlinear responses further provide additional image contrast beyond mean intensity, and improve the visibility and discrimination of target features. By combining flat-band-enhanced frequency conversion with displacement-resolved contrast, this approach expands the information accessible through THG imaging and provides a route to efficient, high-contrast nonlinear image formation.

\section{DESIGN AND SIMULATION}
\noindent
\begin{figure*} [h]
    \centering
    \includegraphics[width=0.76\linewidth]{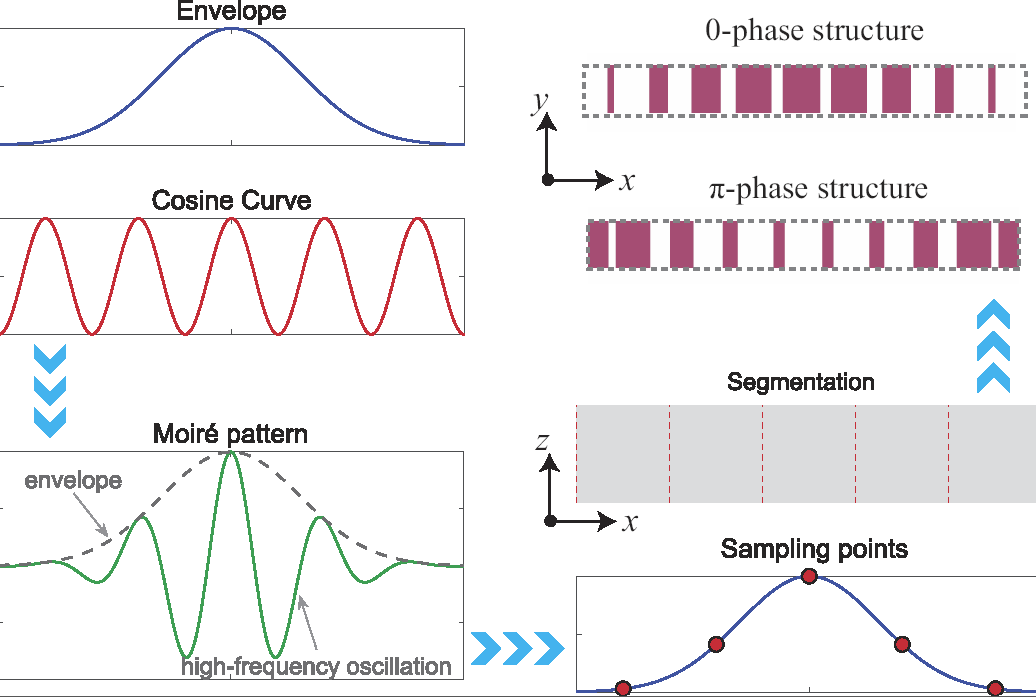}
    \caption{A framework for directly constructing monolayer moir\'e superlattice through an initial moir\'e patterns.
The initial moir\'e pattern is defined by a superposition of a slowly varying envelope and a high-frequency periodic oscillation.  
Monolayer superlattices are generated by segmenting a homogeneous slab into discrete units whose periodicity matches the high-frequency component, with unit widths determined by sampling the moir\'e envelope (red dots).  Two configurations arise depending on the relative phase: the 0-phase and $\pi$-phase structure.}
 \label{Figure_2}
\end{figure*}

A general moir\'e pattern consists of a high-frequency oscillation modulated by a slowly varying envelope. This can be expressed mathematically as
\begin{equation}\label{moir\'eFunct}
P(\mathbf{r}) = A(\mathbf{r}) \cos\left(\mathbf{k}_0 \cdot \mathbf{r} + \phi(\mathbf{r}) \right),
\end{equation}
where \( P(\mathbf{r}) \) is the position-dependent pattern in a single moir\'e period, \( \mathbf{r} \) represents the spatial coordinate, \( A(\mathbf{r}) \) is the moir\'e envelope, and \( \mathbf{k}_0 \) is the spatial vector of the high-frequency oscillation. The \( \phi(\mathbf{r}) \) describes the relative phase between the high-frequency oscillation and the moir\'e envelope at the original point in space. The moir\'e pattern is governed by three key parameters including the spatial vector \(\mathbf{k}_0 \), the moir\'e envelope \( A(\mathbf{r}) \), and the relative phase \( \phi(\mathbf{r}) \). As shown in the first column of Fig.~\ref{Figure_2}, the spatial vector \(\mathbf{k}_0 \) determines the short-range variation of the superlattice, while the moir\'e envelope \( A(\mathbf{r}) \) determines the long-range variation of the superlattice. Finally, the relative phase \( \phi(\mathbf{r}) \) governs the appearance of the moir\'e pattern.

The construction of a general moir\'e superlattice proceeds as follows. 
First, a homogeneous photonic slab is partitioned into discrete units shown in the right-bottom panel in Fig.~\ref{Figure_2}, where the periodicity of segmentation matches the spatial vector of the high-frequency oscillation. 
Then, the width $w$ of each segmented unit is adjusted according to the amplitude of the moir\'e envelope, where the discrete sampling positions are denoted by red dots in Fig.~\ref{Figure_2}.
Consequently, a moir\'e superlattice is achieved with both short-range and long-range variations determined by the moir\'e pattern.
Notably, there are two distinct configurations that arise depending on the segmentation of the high-index homogeneous slab, as shown in the right-top panel in Fig.~\ref{Figure_2}:
(i) \( \phi(\mathbf{r}) = 0 \) leads to a 0-phase structure with the moir\'e envelope aligning with the central high-index segmentation. 
(ii) \( \phi(\mathbf{r}) = \pi \) leads to a $\pi$-phase structure with the moir\'e envelope aligning with the central low-index segmentation.
We note that the relative phase \( \phi(\mathbf{r}) \) could have other values, but here we focus on the two extreme conditions for the sake of simplicity.

In this work, we consider a one-dimensional moir\'e superlattice generated with a single-peak envelope, which is a cosine-type envelope expressed as
\begin{equation}\label{envelope}
A(x) = A_0 + A_1\cos\left(\pi \frac{x}{P_s}\right)^q \quad \text{with} \quad x\in [-P_s/2, P_s/2]\,.
\end{equation}
Here $A_0$ is a constant offset, while $A_1$ is the modulation depth. 
\( P_s \) is the cell length of moir\'e superlattice, and \( q \) is a parameter that modulates the shape of the moir\'e envelope. If \( q < 1 \), the envelope is referred to as a sub-modulated moir\'e envelope, while if \( q > 1 \), the envelope is termed a super-modulated moir\'e envelope.
Such modulation allows for fine-tuning of the moir\'e envelope, which finally affects the structure of the moir\'e superlattice. For a detailed analysis of how these structural parameters influence the moir\'e flatbands, see Section~S1 of the Supporting Information: Tailoring moir\'e flatbands by structural parameters.

\begin{figure*}[htp]
    \centering
    \includegraphics[width=1.0\linewidth]{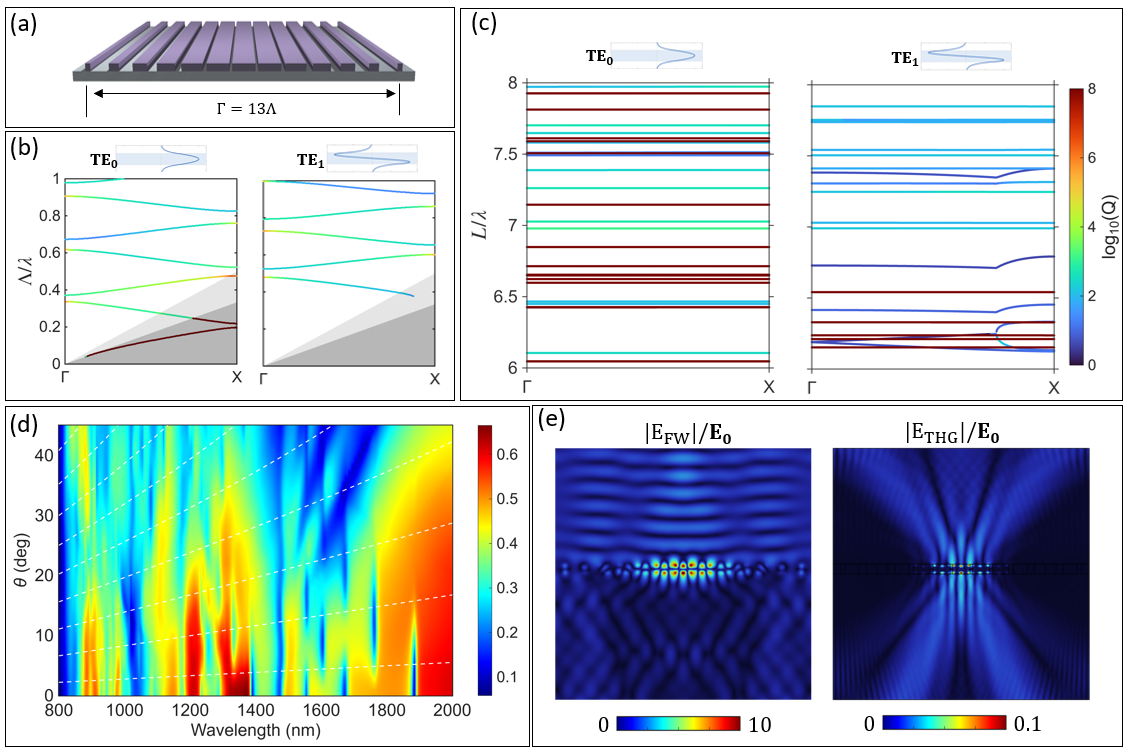}
    \caption{(a) Schematic of the one-dimensional moir\'e superlattice with supercell period $\Gamma=13\Lambda$. (b) Calculated band structures of the conventional periodic grating for the $\mathrm{TE_{0}}$ and $\mathrm{TE_{1}}$ mode families. Grey shading marks the regions below the light lines in the air and substrate sides. (c) Corresponding band structure for the moire\'e superlattice. Colours indicate $\log_{10}{Q}$, where Q is the radiative quality factor of the mode. The inset of (b) and (c) illustrates the reference-slab electric-field profiles along the thickness direction for $\mathrm{TE_{0}}$ and $\mathrm{TE_{1}}$ with zero and one node, respectively. (d) Simulated transmission map as a function of input wavelength and incident angle under y-polarized pump input. The white dashed curves indicate the Brillouin-zone boundaries, defined by $k_{\parallel}=(m+\tfrac{1}{2})2\pi/\Gamma$ where $m$ is an integer. (e) Simulated electric near-field distributions for pump and THG emission for pump at the wavelength near the resonance at  1390 nm.}
    \label{fig:SIM}
\end{figure*}

Following the design principle described above, we engineer a one-dimensional photonic lattice supporting multiple flatbands near (1500~nm) for infrared imaging applications. The grating comprises 13 elements arranged symmetrically about the centre with a uniform pitch of \(\Lambda=800~\mathrm{nm}\) and a height of \(h = 400  ~\mathrm{nm}\). The element widths are modulated by a slowly varying cosine envelope with a superperiod of \(\Gamma=13\Lambda=10.4~\mu\mathrm{m}\). Starting from a nominal width of \(w_0=0.5P=350~\mathrm{nm}\), the width of the \(m\)-th element is defined as \(w_m=w_0+\Delta w[2\cos(\pi x_m/P_{\mathrm{s}})-1]\), where \(x_m=mP\), \(m=-6,\ldots,6\), and \(\Delta w=0.8w_0=280~\mathrm{nm}\). This produces a symmetric width profile ranging from approximately \(138\) to \(630~\mathrm{nm}\) across the 13-element supercell, thereby introducing a long-range spatial modulation into the underlying periodic grating, as shown in Fig.~\ref{fig:SIM}(a).


Figure~\ref{fig:SIM}(a) illustrates the supercell geometry of the designed silicon moir\'e superlattice.
We then calculate the band structures using the guided-mode expansion method implemented in Legume~\cite{minkov2020inverse,zanotti2024legume}. Fig.~\ref{fig:SIM}(b) gives the calculated band structures of a conventional periodic grating for the TE$_0$ and TE$_1$ mode families, i.e., for the case when \(\Delta w=0~\mathrm{nm}\), with representative field profiles along $z$ direction shown in the insets, corresponding to the number of zero nodes of $E_y$ components within the slab. These bands exhibit a strong dispersion along the $\Gamma$ to X direction as expected.

In contrast, the calculated photonic band structure of TE-like modes in Fig.~\ref{fig:SIM}(c) shows multiple weakly dispersive bands throughout the near-infrared spectral range, arising from band folding and modal hybridization within the moir\'e superlattice. These flat-band modes provide multiple resonant channels for light–matter interaction while maintaining their resonance frequencies over a broad range of in-plane wavevectors. The comparison between Figs.~\ref{fig:SIM}(b) and~\ref{fig:SIM}(c) shows that the slowly varying moir'e modulation reshapes the photonic dispersion, producing modes whose resonance frequencies vary weakly with the in-plane wavevector. This reduced dispersion provides a basis for maintaining resonant excitation over a broader range of incident angles.  It is worth mentioning that similar band-flattening effects can also be realized using disorder~\cite{de1989transverse,schwartz2007transport}, such as in disordered photonic lattices, where the relevant bands describe the propagation constant as a function of the transverse wavevector at a fixed optical frequency. The form of disorder-induced long-range superlattice modulation opens minigaps and produces nearly flat minibands, suppressing the transverse beam transport to realize light localization and confinement~\cite{xu2013anomalous}. 

This optical properties of moir\'e superlattice is further confirmed by the calculated angle-resolved transmission map under $y$-polarized plane-wave excitation in Fig.~\ref{fig:SIM}(d), where the spectral position of the corresponding resonances remains nearly unchanged with increasing angle of incidence, demonstrating weak angular dispersion and a broad angular acceptance for mode excitation. 
It is worth mentioning that the one-dimensional moir\'e superlattice exhibits polarization-dependent band flattening. The TE-like bands display substantially weaker dispersion over the investigated spectral range as compared to the TM-like bands. The calculated TM-like band structure for $\mathrm{TM_0}$ and $\mathrm{TM_{1}}$ can be seen in Section 2 of the Supporting Information. This difference reflects the polarization-dependent response of the optical modes to the spatially varying bar widths: dielectric-boundary variations perturb modes differently according to their tangential electric and normal displacement fields~\cite{johnson2002perturbation}.  The flatter TE-like bands are consistent with stronger moir\'e-induced localization and weaker effective coupling between neighbouring supercells, although confirming this interpretation requires analysis of the corresponding eigenfield distributions. A similar connection between stronger localization within a supercell and reduced bandwidth has been reported for TE-like modes in one-dimensional silicon moir\'e photonic crystal slabs~\cite{qin2024optical}. With the grating bars extending along \(y\), these TE-like resonances are preferentially addressed by \(y\)-polarized illumination.

Fig.~\ref{fig:SIM}(e) gives the calculated electric near-field distribution for a moir\'e resonance near 1400~nm, showing pronounced field localization and enhancement within the silicon grating elements. This enhanced pump field further strengthens the third-order nonlinear polarization that drvives the THG process, as illustrated by the third-harmonic electric near-field distribution in the right column of Fig.~\ref{fig:SIM}(e), calculated at an incident pump beam intensity of 1~$\mathrm{GW/cm^2}$. Together with the weakly dispersive bands and angle-resolved optical response, these results suggest the potential of the moiré superlattice to combine resonant nonlinear enhancement with a broad angular response, offering a promising platform for efficient nonlinear imaging over an extended range of incident angles and spatial frequencies.

\begin{figure*} [htp]
    \centering
    \includegraphics[width=0.95\linewidth]{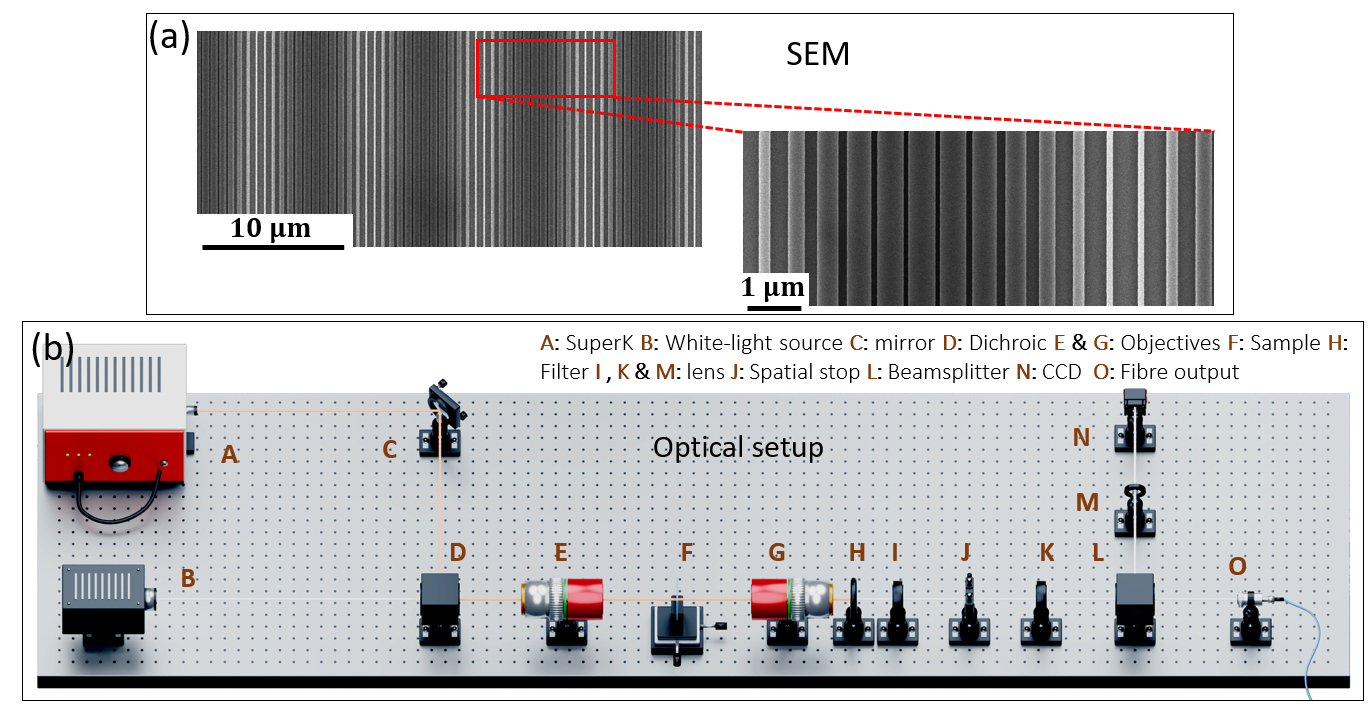}
    \caption{ (a) SEM of the fabricated Moir\'e lattice sample. (b) Experimental setup for the linear transmission measurements and nonlinear optical characterisations.}
 \label{Fig:Setup}
\end{figure*}

\section{EXPERIMENT}
\noindent

The moir\'e superlattice sample was fabricated using the conventional electron-beam lithography (EBL) and inductively coupled plasma (ICP) etching techniques. Figure~\ref{Fig:Setup}(a) shows the scanning electron microscopy (SEM) image of the fabricated sample. As can be seen, the one-dimensional moir\'e superlattice comprises periodically arranged grating elements with systematically modulated widths. The fabricated structures closely reproduce the designed geometry and exhibit well-defined edges and good uniformity over the measured area.
The linear and nonlinear optical responses of the fabricated moir\'e superlattice were characterized using the experimental setup illustrated in Fig.~\ref{Fig:Setup}(b). 
For linear transmission measurements, broadband light from a white-light source was focused onto the sample using a lens with a focal length of 7.5~cm (AC254-075-AB-ML). The transmitted light was collected by a second objective with an NA of 0.4 and coupled into a spectrometer for spectral analysis. 
For nonlinear characterization, we used a broadband supercontinuum laser source (NKT Photonics SuperK FIU-15) to generate an infrared excitation pump beam. A polarizer and a half-wave plate were used to obtain the pump beam with linear polarization along the \(y\)-axis. A longpass dichroic mirror with a cutoff wavelength of 1180~nm (DMLP1180) was used to spectrally filter the pump and to select wavelengths between 1180 and 2000~nm for the excitation. The excitation beam was focused onto the sample through the same lens for THG emission spectra characterization, and through an objective lens with a numerical aperture (NA) of 0.26 for THG imaging characterization, while the generated THG emission was collected in transmission channel using the smae NA = 0.4 objective and directed to the spectrometer and camera.

\begin{figure*} [htp]
    \centering
    \includegraphics[width=0.8\linewidth]{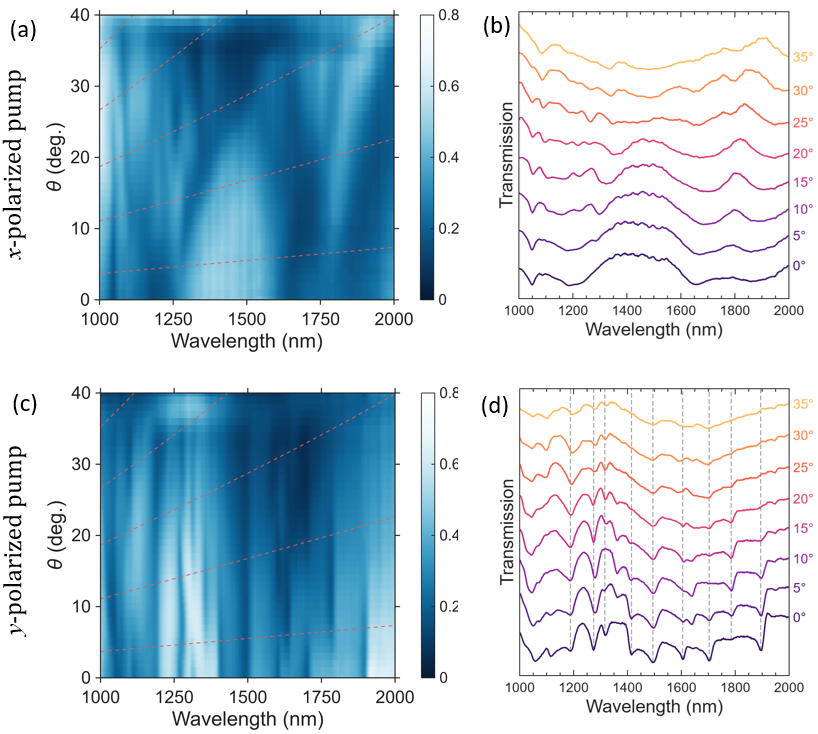}
    \caption{Experimentally measured transmission of the moir\'e lattice sample. (a) Transmission map as a function of wavelength and incidence angle for $x$-polarized illumination. (b) Corresponding transmission spectra at selected incidence angles. (c) Transmission map for $y$-polarized illumination. (d) Corresponding transmission spectra at selected incidence angles. The dashed curves in (a, c) indicate the Brillouin-zone boundaries,defined by $k_{\parallel}=(m+\tfrac{1}{2})2\pi/\Gamma$ where $m$ is an integer.}
 \label{Fig:Trans}
\end{figure*}


The experimentally measured angle-resolved transmission spectra are shown in Fig.~\ref{Fig:Trans}. Figures~\ref{Fig:Trans}(a) and \ref{Fig:Trans}(b) show the measured transmission map and selected spectra under $x$-polarized pump illumination. As can be seen, no distinct excitation feature of moir\'e modes is observed. In contrast, under $y$-polarized pump illumination, we observe pronounced transmission minima in Fig.~\ref{Fig:Trans}(c, d), indicating clear excitation of the moir\'e modes. The transmission spectra reveal multiple resonant bands across the near-infrared range from \(1000\) nm to \(2000~\mathrm{nm}\), consistent with the multiple modes predicted by the calculated band structure in Fig.~\ref{fig:SIM}. In particular, several pronounced transmission minima between approximately \(1200\) and \(1500~\mathrm{nm}\) exhibit only small spectral shifts with increasing angle of incidence, confirming their weak angular dispersion.
Although the flatband resonance wavelengths remain weakly dispersive, their transmission signatures vary with incidence angle, reflecting changes in radiative coupling and interference with the nonresonant transmission background.
The dashed curves mark successive Brillouin-zone boundaries of the moir\'e superlattice. Across these zones, the resonant features exhibit recurrent angular modulation, consistent with repeated sampling of the folded moir\'e bands and angle-dependent radiative coupling~\cite{qin2024optical}.
The spectra in Fig.~\ref{Fig:Trans}(d) further resolve these resonances more clearly and show that their spectral positions remain stable over a broad angular range. We also observe some resonace exhibit angle-dependent broadening and variations in contrast, which may reflect changes in radiative coupling and mode excitation. Despite these variations, their spectral positions remain nearly unchanged over the measured angular range, consistent with the weakly dispersive moir\'e modes. 
This experimentally observed angular robustness provides a basis for nonlinear conversion of multiple spatial-frequency components in nonlinear imaging applications.

Furthermore, varying the lattice pitch $\Lambda$ enables spectral tuning of the moir\'e resonances, with larger and smaller pitches shifting the resonant features towards longer and shorter wavelengths, respectively. This geometric control allows flat-band modes with distinct spatial field profiles to be positioned within targeted spectral regions. For details, see the simulated and experimentally measured transmission spectra for lattice pitches being $\Lambda=600$~nm and $\Lambda=700$~nm for comparison in Figure S6 in the Supporting Information.

\begin{figure*} [h]
    \centering
    \includegraphics[width=0.8\linewidth]{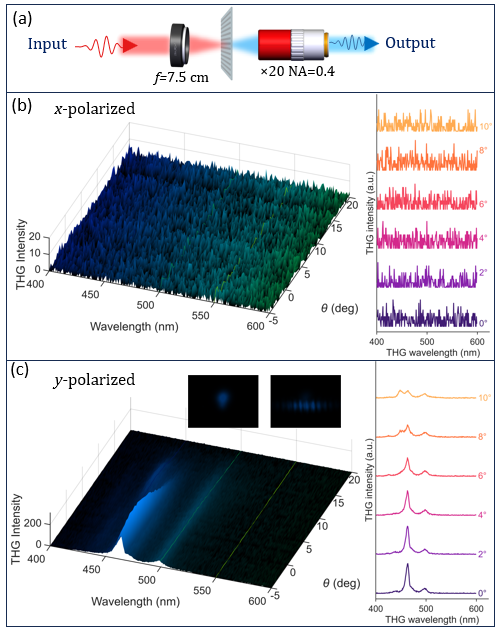}
    \caption{Experimentally measured third harmonic generation (THG). (a) Schematic of the THG measurement setup. (b) THG emission map for $y$-polarized pump excitation. Left: THG intensity map as a function of emission wavelength and pump incidence angle. The inset shows the normalised spatial intensity distribution of THG emission on a silicon film (left) and the superlattice (right) for comparison. Right: THG spectra at selected pump incidence angles. (c) THG emission map for $x$-polarized pump excitation.}
 \label{Fig:THG}
\end{figure*}

The angle-dependent THG response of the fabricated moir\'e superlattice was characterized using the transmission configuration illustrated in Fig.~\ref{Fig:THG}(a).  The pump beam, with a total power of 83~mW integrated over the 1180 - 2000~nm spectral range and measured before the focusing lens, was focused onto the sample using a lens with a focal length of 7.5~cm (AC254-075-AB-ML). This lens produced a relatively large focal spot on the moir\'e superlattice to facilitate excitation of its resonant modes. The frequency-converted visible emission was collected from the opposite side of the sample using a NA=0.4 objective. 
Under $x$-polarized pump excitation, no detectable THG emission was observed from the moir\'e superlattice as shown in Fig.~\ref{Fig:THG}(b), consistent with the absence of resonant mode excitation. Under $y$-polarized broadband pump excitation, the collected angle-resolved THG emission spectra map is shown in the left panel of Fig.~\ref{Fig:THG}(c), a pronounced THG peak is observed near \(\approx\) 467~nm, corresponding to the excitation of a moir\'e mode around a fundamental wavelength of 1400~nm (Fig.~\ref{fig:SIM}(c,d)). Such a moir\'e mode enhances the local electromagnetic field and thereby boosts the nonlinear light–matter interaction. The THG emission peak remains spectrally stable as the pump incidence angle varies, while its intensity exhibits an angular dependence, consistent with the mode excitation observed in the linear transmission spectra map shown in Fig.~\ref{Fig:Trans}. The selected spectra shown in the right panel of Fig.~\ref{Fig:THG}(c) show that the THG peak reamins nearly stable in both wavelength and intensity from  \(0^\circ\) to \(5^\circ\), extending beyond the first Brillouin-zone boundary. At larger angles, such as \(10^\circ\), the peak wavelength remains stable, while the emission intesnity varies as expected.

\begin{figure*} [h]
    \centering
    \includegraphics[width=0.9\linewidth]{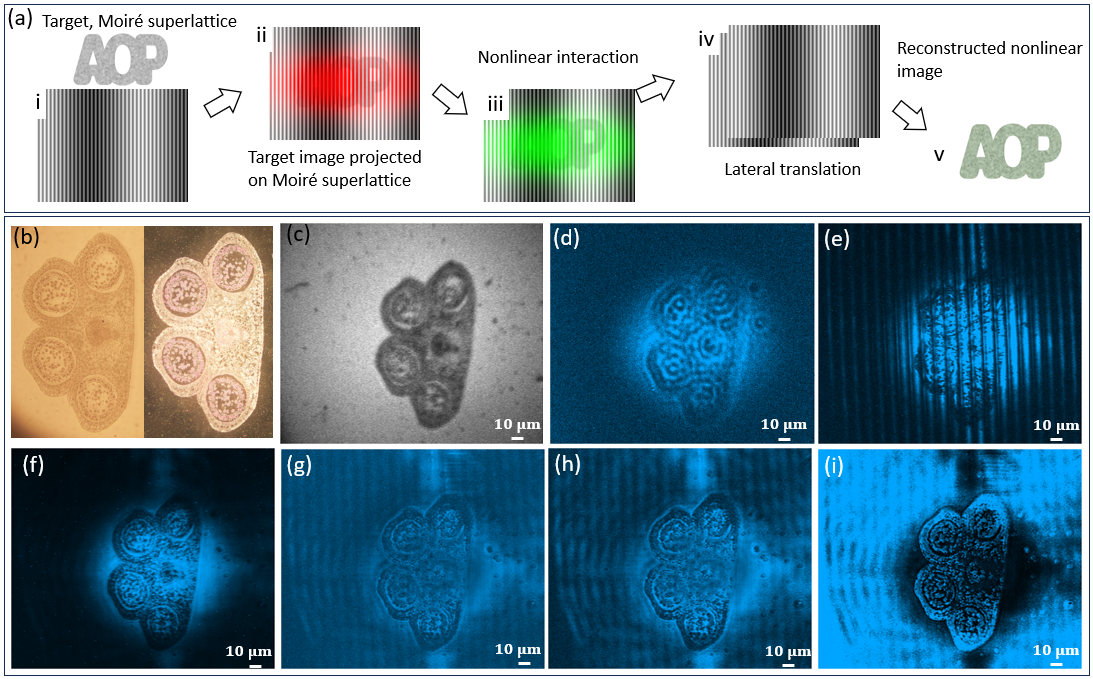}
    \caption{Moir\'e superlattice-assisted THG image processing. (a) Work principle of the imaging process. (b) Brightfield (left) and darkfield (right) optical micrographs of the target. (c) white-light image of the target acquired using our optical setup in Fig.~\ref{Fig:Setup}(b). (d) THG image obtained with a 400-nm thick silicon film. (e-i) THG images with the Moir\'e superlattice: (e) a single THG frame; (f) the average of 30 translated frames; (g) the relative RMS variation map, (h) the harmonic modulation RMS map, and (i) the THG fingerprint decorrelation image. }
 \label{Fig:THGimaging}
\end{figure*}

Finally, we demonstrate a general approach to achieve and improve the THG imaging with the moir\'e superlattice. Figure~\ref{Fig:THGimaging}(a) shows the working principle of moir\'e-lattice-assisted THG imaging. The infrared image of the target is first projected onto the moir\'e lattice, where the local target texture perturbs the excitation of the strongly confined moir\'e modes. Owing to the nonlinear dependence \(I_{\mathrm{THG}}\propto |E|^6\), these local field perturbations are amplified in the generated THG signal. The lattice is then translated laterally in \(1~\mu\mathrm{m}\) increments, and 30 THG images are recorded, with each displacement providing a different spatial encoding of the target. 
Figure~\ref{Fig:THGimaging}(b) shows the target used in the experiment - a lilium flower cell sample, where both the bright-field and dark-field micrographs are presented. 
In our linear imaging experiment, the target is first imaged onto the intermediate imaging plane using a $10\times$ objective with a numerical aperture (NA) of 0.26. The finite NA imposes a spatial-frequency cutoff, such that high-spatial-frequency components of the target are attenuated or rejected during formation of the intermediate image. The resulting bandwidth-limited field is subsequently relayed to the camera using a second objective with a higher NA of 0.40. Although the larger collection NA enables efficient transfer of the intermediate image, it cannot recover spatial information removed by the first imaging stage. Consequently, the image recorded by the camera represents an NA-limited version of the original target, providing a conventional reference against which the additional spatial sensitivity and information encoded through moir\'e-superlattice-mediated THG imaging can be evaluated. A more detailed description about the image formation with the optical setup can be seen in S4: Image formation and nonlinear feature sharpening, of the Supporting Information.
Figure~\ref{Fig:THGimaging}(c) shows the white-light image of the target acquired using our optical setup, which indeed exhibits noticeable blurring and loss of fine spatial features.
For further comparison, Figure~\ref{Fig:THGimaging}(d) gives the corresponding THG image obtained using an unpatterned \(400\)-nm-thick silicon reference film under the same broadband SuperK pump illumination. 
In the following, we experimentally demonstrate that introducing the moir\'e superlattice at the intermediate image plane transforms this NA-limited optical field into displacement-dependent nonlinear mode fingerprints, providing enhanced spatial sensitivity and contrast and enabling improved discrimination of weak spatial features.

Lateral translation of the moir\'e superlattice modifies the spatial overlap between the projected target field and the resonant mode profiles, producing displacement-dependent THG emission. The nonlinear response originates from the third-order polarization~\cite{boyd2008nonlinear},
\begin{equation}
\begin{aligned}
    P_{3\omega,i}(\mathbf{r};\Delta_k)
    &= \varepsilon_0 \sum_{j,\ell,m}
    \chi^{(3)}_{ij\ell m}(\mathbf{r};\Delta_k) \\
    &\quad \times E_{\omega,j}(\mathbf{r};\Delta_k)
    E_{\omega,\ell}(\mathbf{r};\Delta_k)
    E_{\omega,m}(\mathbf{r};\Delta_k).
\end{aligned}
\end{equation}
where $i,j,\ell,m \in \{x,y,z\}$,  $\Delta_k$ denotes the $k$th lattice displacement. The cubic dependence on the local excitation field makes the nonlinear emission sensitive to displacement-induced changes in resonant excitation, while the recorded images also depend on coherent propagation and collection of the generated field. The moire\'e flat-band modes combine weak angular dispersion with spatially nonuniform field profiles within the supercell. These properties play complementary roles: weak dispersion supports resonant excitation of different transverse-wavevector components. At the same time, the structured field profiles make their overlap with the projected target field sensitive to lattice displacement. The obtained THG variations therefore contain information that is lost upon averaging. We aim to extract this information in the following.

To characterize this response, we analyze the measured THG image sequence $I_k(\mathbf{R})$, where $k$ indexes the lattice displacement and $\mathbf{R}$ denotes a camera coordinate, with all frames sharing the same target coordinate system. Subsequently, we define the mean image and relative root-mean-square (RMS) variation as
\begin{equation}
\label{eq:Vrms}
\begin{aligned}
    \overline{I}(\mathbf{R})
    &= \frac{1}{N}\sum_{k=1}^{N} I_k(\mathbf{R}), \\[6pt]
    V_{\mathrm{RMS}}(\mathbf{R})
    &= \frac{
        \sqrt{\frac{1}{N}\sum_{k=1}^{N}
        [I_k(\mathbf{R})-\overline{I}(\mathbf{R})]^2}
    }{
        \max[\overline{I}(\mathbf{R}),\tau]
    }.
\end{aligned}
\end{equation}
where $N$ is the number of displacement positions and $\tau$ is an intensity floor that stabilises the normalisation to prevent division by zero or extremely weak signals. 
The mean image captures the average THG intensity, whereas the relative RMS variation measures the amplitude of displacement-induced intensity fluctuations relative to the local mean. This normalization highlights regions whose nonlinear emission is particularly sensitive to the lattice translation, rather than simply those with high intensity, revealing spatial features that may be less apparent in the mean intensity image. We further introduce the harmonic modulation RMS to further reduce the sensitivity to fluctuations in the background noise, see details in S5: Displacement-dependent THG contrast analysis of the Supporting Information, 
To identify spatial differences in the scan response, each pixel is assigned a mean-subtracted fingerprint that records how its THG intensity varies throughout the scan,
$\mathbf{f}(\mathbf{R})=[I_1(\mathbf{R})-\overline{I}(\mathbf{R}),\ldots,I_N(\mathbf{R})-\overline{I}(\mathbf{R})]^{\mathrm{T}}$.
A local decorrelation metric $D$ compares this fingerprint with those of neighbouring pixels, highlighting regions where nearby pixels respond differently to the lattice translation, even when their mean intensities are similar. 
The local fingerprint decorrelation is then defined as
\begin{equation}\label{eg:DR}
    D(\mathbf{R})
    =
    \frac{1}{|\mathcal{N}_{\mathbf{R}}|}
    \sum_{\mathbf{R}'\in\mathcal{N}_{\mathbf{R}}}
    \left[
    1-
    \frac{\mathbf{f}(\mathbf{R})^{\mathrm{T}}\mathbf{f}(\mathbf{R}')}
    {\|\mathbf{f}(\mathbf{R})\|_2\|\mathbf{f}(\mathbf{R}')\|_2}
    \right],
\end{equation}
where $\mathcal{N}_{\mathbf{R}}$ contains the valid immediate horizontal and vertical neighbours of pixel $R$. Here, weak-signal pixels are masked, and fingerprints with vanishing norms are excluded. Whereas $V_{\mathrm{RMS}}$ highlights regions with strong displacement-induced intensity variations relative to their mean brightness - revealing weak features that have little contrast in the mean image, $D$ identifies neighbouring pixels with distinct responses, indicating internal contours and fine textures that have similar average brightness. Together, these complementary measures provide spatial contrast beyond the mean THG brightness. Full analysis details are given in S5: Displacement-dependent THG contrast analysis, of the Supporting Information.

Figure~\ref{Fig:THGimaging}(e-i) gives the nonlinear imaging results obtained by laterally translating the moir\'e lattice relative to the intermediate target image. Figure~\ref{Fig:THGimaging}(e) shows a single-frame THG image at one lattice position, where the target information is encoded through the spatially dependent excitation of the moir\'e modes and the resulting nonlinear emission. Figure~\ref{Fig:THGimaging}(f) shows the mean intensity THG image $\overline{I}(\mathbf{R})$ averaged over all displacement positions, revealing the overall target profile while suppressing position-dependent intensity variations due to the superlattice structure. To extract additional information carried by these variations, we further analyze the displacement-dependent image sequence using a computational imaging technique that accounts for the optical characteristics of the moir\'e-mode excitation, as mentioned above. 
Figure~\ref{Fig:THGimaging}(g) and Figure~\ref{Fig:THGimaging}(h) shows the relative RMS variation (Eq.~\ref{eq:Vrms}), quantifying the normalised strength of the displacement-dependent THG modulation, and the harmonic modulation by considering up to 4th order harmonics, based on Eq.~S31 of the Supporting Information, to capture the fitted periodic modulation up to the fourth harmonic and reduced contributions from noise outside the selected harmonics. Fig.~\ref{Fig:THGimaging}(i) shows the local fingerprint decorrelation map, defined by Eq.~\ref{eg:DR}, which highlights spatial regions exhibiting distinct changes in their nonlinear response during translation. Compared with the bright-field and dark-field optical micrographs, as well as the white-light image acquired through our setup, the displacement-dependent moir\'e THG measurements exhibit substantially enhanced spatial sensitivity and contrast. In particular, nonlinear imaging using the nonlinear moir\'e superlattice resolves weak spatial features that are blurred or difficult to distinguish from conventional nonlinear image techniques. More examples can be seen in S6: Nonlinear imaging with moir\'e superlattices, in the Supporting Information. Together, these measurements demonstrate that the spatially structrued resonant response of the moir\'e superlattice is able to encode image information into the displacement-dependent THG emission pattern, revealing weak features and internal structure that are less apparent in conventional nonlinear intensity images, including single-frame or mean-intensity THG images.

\section{Conclusions}
In conclusion, we have demonstrated a photonic moir\'e-superlattice platform that combines flat-band-enhanced THG with nonlinear spatial encoding and imaging. The engineered one-dimensional moir\'e superlattice supports multiple quasi-flat bands in the near-infrared spectral region, providing strong field confinement together with broad momentum-space acceptance. Angle-resolved linear and nonlinear measurements confirm spectrally stable moir\'e mode excitation and robust THG generation over a finite angular range, alleviating the conventional trade-off between resonant field enhancement and angular acceptance. Beyond enhanced nonlinear frequency conversion, the moir\'e superlattice provides a spatially structured nonlinear response that can be employed for imaging. When placed at the intermediate image plane, the lattice interacts with an NA-limited representation of the target, in which fine spatial features are partially attenuated by the preceding imaging optics. Lateral translation of the moir\'e lattice continuously changes the spatial overlap between the target field and the enhanced moir\'e mode profiles, modifying their local excitation and generating a sequence of displacement-dependent nonlinear fingerprints. The strong intensity dependence of THG further amplifies these variations, making weak spatial differences more distinguishable in the nonlinear response. Analysis of the displacement-dependent image sequence through relative RMS variation and local fingerprint decorrelation reveals spatial features and internal textures with enhanced sensitivity and contrast, as compared with conventional bright-field, dark-field, and white-light imaging results. These results demonstrate that the moir\'e superlattice not only can serve as an efficient nonlinear frequency converter with a robust angular response, but also can serve as a spatial encoder that maps the overlap between the incident field and resonant mode profiles into displacement-dependent THG signals. Lattice translation thus adds an additional measurement dimension beyond intensity alone, enabling improved discrimination of target features. 
With further optimization of the resonance field profiles, spatial sampling, and reconstruction algorithms, we expect this approach could further extend its capability towards label-free, super-resolution nonlinear imaging.

\subsection*{Conflicts of Interest}

The authors declare no conflicts of interest.

\subsection*{Supporting Information}
Supporting Information is available from the authors.

\bibliographystyle{unsrt}
\bibliography{rfs}

\end{document}